\documentclass[submission, Proceedings]{SciPost}
\hypersetup{
    colorlinks,
    linkcolor={red!50!black},
    citecolor={blue!50!black},
    urlcolor={blue!80!black}
}

\usepackage[bitstream-charter]{mathdesign}
\usepackage{rotating}
\usepackage{lscape}

\DeclareSymbolFont{usualmathcal}{OMS}{cmsy}{m}{n}
\DeclareSymbolFontAlphabet{\mathcal}{usualmathcal}
\def\apj{ApJ}%
\definecolor{cyan}{cmyk}{1.,0.,0.,0.2}
\definecolor{vert}{cmyk}{0.5,0.,0.5,0.5}
\definecolor{magenta}{cmyk}{0.,1.,0.,0.1}
\definecolor{verdatre}{cmyk}{0.5,0.,0.5,0.5}
\definecolor{vert_clair}{cmyk}{0.5,0.,0.5,0.2}
\definecolor{yellow}{cmyk}{0.,0.,1.,0.0}
\definecolor{yellow_1}{cmyk}{0.,0.,0.5,0.0}
\definecolor{rouge}{cmyk}{0.,0.4,0.6,0.0}
\definecolor{orange}{cmyk}{0.,0.5,0.5,0.05}
\definecolor{violet}{rgb}{0.5,0.,0.5}
\definecolor{darwin_box}{rgb}{0.988,0.878,0.77}
\definecolor{darwin_text}{rgb}{0.1,0.07,0.02}
\definecolor{blue_mertsch}{rgb}{0.08,0.05,0.25}
\definecolor{blue_light_mertsch}{rgb}{0.7,0.67,0.77}
\newcommand{\mC}{\mbox{$m_{\chi}$}}
\newcommand{\sigv}{\langle \sigma v\rangle}
\def\big{\textsf{BIG}}
\def\slim{\textsf{SLIM}}
\def\quaint{\textsf{QUAINT}}
\begin{document}
\begin{center}{\Large \textbf{
What charged cosmic rays tell us on dark matter\\
}}\end{center}

\begin{center}
Pierre Salati\textsuperscript{$\,\star$}
\end{center}

\begin{center}
LAPTh and Universit\'e Savoie Mont-Blanc\\
9 chemin de Bellevue, BP110, Annecy-le-Vieux\\
F-74941 Annecy Cedex, France
\\
* pierre.salati@lapth.cnrs.fr
\end{center}

\begin{center}
LAPTH-Conf-065/22
\end{center}


\definecolor{palegray}{gray}{0.95}
\begin{center}
\colorbox{palegray}{
  \begin{tabular}{rr}
  \begin{minipage}{0.1\textwidth}
    \includegraphics[width=30mm]{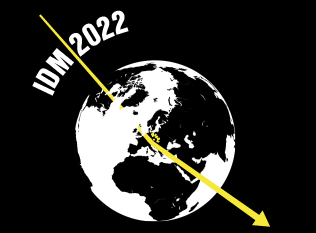}
  \end{minipage}
  &
  \begin{minipage}{0.85\textwidth}
    \begin{center}
    {\it Plenary talk given at the}\\
    {\it 14th International Conference on Identification of Dark Matter}\\
    {\it Vienna, Austria, 18-22 July 2022} \\
    \end{center}
  \end{minipage}
\end{tabular}
}
\end{center}
\section*{Abstract}
{\bf
%
Dark matter particles could be the major component of the haloes of galaxies. Their mutual annihilations or decays would produce an indirect signature under the form of high-energy cosmic-rays.
The focus of this presentation is on antimatter species, a component so rare that any excess over the background should be easily detected. After a recap on Galactic propagation, I will discuss positrons, antiprotons and anti-nuclei. For each of these species, anomalies have been reported. The antiproton excess, for instance, is currently a hot topic. Alas, it does not resist a correct treatment of theoretical and data errors.
}


\section{Introduction}
\label{sec:intro}

A significant portion of the universe is in the form of a massive and pressureless component, dubbed dark matter (DM). Nearly 90 years after DM was discovered by Zwicky in the Coma galactic cluster~\cite{1937ApJ....86..217Z}, its nature is still unresolved. Among the plethora of candidates so far proposed, weakly interacting, massive and neutral species have attracted much attention. These particles would pervade the DM haloes of galaxies inside which they are expected to annihilate or decay, producing gamma-rays, neutrinos and charged particles, in particular antimatter cosmic rays (CRs) which are the focus of this contribution.

Positrons, antiprotons and anti-nuclei are already produced by high-energy CR protons and helium nuclei interacting on interstellar gas, hence yielding backgrounds against which a potential DM anomaly is to be searched under the form of a spectral excess. Most of these backgrounds have been detected. Understanding them is the keystone to correctly decipher and validate a DM signal. This requires to model accurately the propagation of charged CRs inside the turbulent magnetic fields of the Milky Way. These extend over a region, the so-called magnetic halo, which encompasses the Galactic disk and extends above and beneath it over a distance $L$. The thicker the magnetic halo, the more DM is trapped inside it and the stronger the CR flux at the Earth from DM origin. Determining this height $L$ is paramount. It requires to properly model the propagation of charged species inside the Milky Way.

\section{Measuring the height of the magnetic halo}
\label{sec:height_halo}

Charged particles follow the Galactic magnetic lines and diffuse on their knots. CR transport is actually described as a diffusion process in space, with coefficient $K$ grossly scaling with rigidity as ${\cal R}^{\delta}$. A spectral break of $K$ has been observed at high rigidity, making CR fluxes harder above ${\cal O}(250)$ GV \cite{Genolini:2017dfb}. At rigidities of order a few GV, the situation is unclear. Either diffusion becomes more efficient at low energy, hence another spectral break, or Galactic convection as well as diffusive reacceleration come into play. Both possibilities are allowed in the transport scheme {\big} proposed in \cite{Genolini:2019ewc} to fit the secondary-to-primary flux ratio B/C, a sensitive probe of CR diffusion insofar as it scales like $L/K$. The {\quaint} and {\slim} schemes are two subsets of {\big}. The former corresponds to the configurations used so far in the literature while the latter is devised for the gifted amateur.

Measuring the flux of an unstable secondary nuclear species is necessary to break the degeneracy between the diffusion coefficient $K$ and the height $L$ of the magnetic halo. Beryllium is a secondary species produced, like boron, by the fragmentation of carbon nuclei impinging on interstellar gas. Its unstable isotope $^{10}$Be has a half-lifetime $t_{1/2}$ of 1.387 Myr of the same order of magnitude as the typical resident time of CRs inside the Galactic disk. At high energy, $^{10}$Be behaves as if it were stable. At low energy, it propagates over a distance $\lambda \propto (K t_{1/2})^{1/2}$
and its flux at the Earth scales like ${1}/{\lambda}$, hence the possibility to disentangle $K$ from $L$.

To do so requires the flux ratios {$^{10}$Be}/{Be} and {$^{10}$Be}/{$^{9}$Be}. However, these have been essentially measured below 1 GeV/n and suffer from lack of statistics. On the other hand, some isotopic information is contained in the elemental ratio {Be}/{B} recently measured by AMS-02 with great precision \cite{PhysRevLett.120.021101}. Combining all data sets yields a halo size $L$ of order $4.5 \pm 1$ kpc as showed in~\cite{Weinrich:2020ftb}. The best-fit value $L$ depends actually on the CR scheme and is respectively equal to $4.08$ ({\quaint}), $4.64$ ({\big}) and $4.66$~kpc ({\slim}).

\section{Positrons, dark matter and TeV haloes}
\label{sec:positrons}

In 2008, the rise of the positron fraction reported by the PAMELA collaboration \cite{PAMELA:2008gwm} triggered a hectic activity to explain the observed excess in terms of DM annihilations or decays. The dust has settled by now. On top of the background of secondary positrons produced in the Galactic disk by primary CR nuclei interacting with gas, a primary component is also contributed by nearby pulsars as already proposed in 1988 by A.~Boulares~\cite{1989ApJ...342..807B}, hence a positron flux
\begin{equation}
\Phi_{e^{+}} = \Phi_{e^{+}}^{\rm sec} + \Phi_{e^{+}}^{\rm prim}({\rm pulsars}) \,.
\end{equation}
The primary component is underdetermined since a single pulsar is enough to fit the observations~\cite{2013ApJ...772...18L,Boudaud:2014dta}. Research is currently focused on the kind of pulsar population needed to fit the data. As showed in the comprehensive simulation carried out in~\cite{Orusa:2021tts}, most of the synthetic populations of pulsar wind nebulae (PWNe) produce several wiggles in the positron flux which are not observed. The possibility of just a few sources, typically around 3, generating a flux over a wide range in energy is then favored. This result is criticized in~\cite{John:2022asa} where the stochastic nature of high-energy inverse Compton scatterings (ICS), a prominent cooling mechanism for positrons, is showed to erase the wiggles in the flux sourced by PWNe. This reopens the possibility that such a feature could be used as a signature for DM.

In this context, it is somewhat difficult to extract robust bounds on the annihilation cross section or decay rate of DM species. A conservative approach would consist in disregarding pulsars and adding to the sole secondary positron component the flux produced by DM, requiring that none of the data points is overshooted. This yields upper bounds on $\sigv$ above the thermal value of $3 \times 10^{-26} \, {\rm cm^{3} \, s^{-1}}$ required to get the observed DM cosmological abundance \cite{DiMauro:2021qcf}.
A more agressive possibility is to fit the data and to gauge how much DM can be added without perturbing the fit. This yields considerably more stringent bounds which, in the case of the $e^{+} e^{-}$ channel, lie below the thermal value for DM masses up to 200 GeV \cite{DiMauro:2015jxa,DiMauro:2021qcf}.

The recent discovery of $\gamma$-ray TeV haloes around pulsars \cite{HAWC:2017kbo,DiMauro:2019yvh} confirms these objects as potential sources of positrons. These are accelerated by pulsar winds and produce TeV photons through ICS on stellar light inside a region where space diffusion is significantly inhibited \cite{Manconi:2020ipm} compared to the bulk of the CR magnetic halo. Current studies concentrate on how magnetic turbulence is itself generated by the accelerated particles \cite{Mukhopadhyay:2021dyh}.

%
\renewcommand{\arraystretch}{1.3}
\begin{sidewaystable}
\centering
\begin{tabular}{||l||c|c|c|c|c|c||}
\hline
Authors & Data analyzed & Method & ${\cal C}^{\rm data}$ & ${\cal C}^{\rm model}$ & Significance of & Conclusive \\
 &   & F or B & & & fit w and w/o DM & hint for DM \\
\hline
\hline
Cuoco et al. \cite{Cuoco:2016eej} & $\bar{p}/p$, $p$ \& He + Voyager $p$ \& He & F & no & no & $4.7\,\sigma$ (L,2dof) & \textcolor{verdatre}{yes} \\
\hline
Cui et al. \cite{Cui:2016ppb} & $\bar{p}$ with prior $\Pi(\lambda)$ from CR nuclei & B & no & no & $2_{\,}\ln\!{K} = 11$ to $54$ & \textcolor{verdatre}{Decisive} \\
\hline
Reinert \& Winkler \cite{Reinert:2017aga} & B/C, $\bar{p}$ \& AMS/PAM & F & no & ${\cal C}^{\rm XS , parents}$ & $2.2\,\sigma$ (L,1dof) $\to$ $1.1\,\sigma$ (G) & \textcolor{red}{no} \\
\hline
Cui et al. \cite{Cui:2018nlm} & $\bar{p}$ with prior $\Pi(\lambda)$ from CR nuclei & B & no & no & Best-fit $\mC$ and $\sigv$ & N/A \\
 & + PandaX-II, LUX, XENON1T (DD) & & & & & \\
\hline
Cuoco et al. \cite{Cuoco:2019kuu} & $\bar{p}/p$, $p$ \& He + Voyager and CREAM $p$ \& He & F & no & no & $3.1\,\sigma$ (L,2dof) & \textcolor{verdatre}{yes} \\
 & all data with ${\cal R} > 5 \, {\rm GV}$ & F & no & ${\cal C}^{\rm XS}$ & $2.9\,\sigma$ (L,2dof) & \textcolor{verdatre}{yes} \\
 & & F & {\checkmark} & no & $5.5\,\sigma$ (L,2dof) & \textcolor{verdatre}{yes} \\
\hline
Cholis et al. \cite{Cholis:2019ejx} & $\bar{p}/p$ with best-fit of fudge factor $N_{XS}(T_{\bar{p}})$ & F & no & no & $4.7\,\sigma$ (L,1dof) & \textcolor{verdatre}{yes} \\
 & + $\bar{p}$ production in SNR shocks & F & no & no & $3.3\,\sigma$ (L,1dof) & \textcolor{verdatre}{yes} \\
\hline
Heisig et al. \cite{Heisig:2020nse} & B/C, $\bar{p}$ \& AMS/PAM & F & {\checkmark} & ${\cal C}^{\rm XS}$ & $0.5\,\sigma$ (L,1dof) & \textcolor{red}{no} \\
 & $\bar{p}/p$, $p$ \& He + Voyager $p$ \& He & F & {\checkmark} & ${\cal C}^{\rm XS}$ & $1.8\,\sigma$ (L,1dof) $\to$ $0.5\,\sigma$ (G) & \textcolor{red}{no} \\
\hline
Di Mauro \& Winkler \cite{DiMauro:2021qcf} & B/C, $\bar{p}$ \& AMS/PAM & F & {\checkmark} & ${\cal C}^{\rm XS}$ & $2\,\sigma$ (L,1dof) $\to$ $1\,\sigma$ (G) & \textcolor{red}{no} \\
\hline
Luque \cite{Luque:2021ddh} & B/C, B/O, Be/C, Be/O, Be/B, Li/B, Li/Be & B & no & no & DM fit is poorer & \textcolor{red}{no} \\
 & + $^{10}$Be/Be, $^{10}$Be/$^{9}$B and $\bar{p}/p$ & & & & than fit w/o DM& \\
\hline
Kahlhoefer et al. \cite{Kahlhoefer:2021sha} & $p$, He \& $\bar{p}/p$ + Voyager $p$ \& He & F+B & no & no &  $4.5\,\sigma$ (L,1dof) & \textcolor{orange}{TBC} \\
\hline
Calore et al. \cite{Calore:2022stf} & $\bar{p}$ + nuisance on $L$ from LiBeB & F & {\checkmark} & {\checkmark} & $1.8\,\sigma$ (L,1dof) & \textcolor{red}{no} \\
\hline
\hline
\end{tabular}
\caption{Recent analyses of a potential DM signal in the CR $\bar{p}$ flux. The statistical method is either frequentist (F) or Bayesian (B). It makes use ($\checkmark$) or not (no) of a covariance matrix of errors for data (${\cal C}^{\rm data}$) and theory (${\cal C}^{\rm model}$). The significance of the DM signal is either local (L) or global (G).}
\end{sidewaystable}
%
\section{Antiprotons -- Trimmed hints and robust bounds}
\label{sec:antiprotons}

A DM excess hidden in the $\bar{p}$ flux has been claimed by many groups in the past few years. Some studies reach the opposite conclusion. It is timely then to analyze how a DM $\bar{p}$ signal could be extracted from the data and to understand the crucial differences between these analyses which are summarized in Table~1.
A powerful estimator for testing the null-hypothesis, i.e. the absence of any DM signal from observations, is the likelihood ratio
\begin{equation}
LR({\rm null}) = - 2 \ln \left\{
{\displaystyle \frac{{\rm sup}_{\lambda \in \Lambda}{\cal L}(\lambda)}{{\rm sup}_{{\{\lambda,\mu\}} \in \Lambda \cup M}{\cal L}(\lambda,\mu)}}
\right\} \,,
\end{equation}
where ${\lambda \in \Lambda}$ stands for the CR parameters while $\mu \equiv \{ \mC , \sigv , {\rm channel} \}$ characterizes the properties of the DM species. This ratio gauges how more likely the data were to be collected when DM is added.

Defining the likelihood function ${\cal L}(\lambda,\mu)$ requires some care though. In principle, we could define it through a global $\chi^{2}$ measuring the distance of both CR nuclear species and antiproton fluxes to theory
\begin{equation}
- 2 \ln {\cal L}(\lambda,\mu) \equiv \chi_{\rm LiBeB}^{2}(\lambda) + \chi_{\bar{p}}^{2}(\lambda,\mu) \,.
\label{eq:definition_ln_cal_L_1}
\end{equation}
However this would be extremely resources and time consuming.
The situation considerably simplifies by remarking that the $\bar{p}$ flux, which may be expressed as the sum
\begin{equation}
\Phi_{\bar{p}} = \Phi_{\bar{p}}^{\rm sec}(\lambda) + \Phi_{\bar{p}}^{\rm DM}(\lambda , \mu) \,,
\end{equation}
is by far dominated by the contribution from secondaries produced by CR protons and He nuclei interacting with Galactic gas. This component behaves like a stable secondary nuclear species. It is no suprise then if the CR parameters $\hat{\lambda}$ minimizing $\chi^{2}_{\rm LiBeB}$ should also minimize $\chi^{2}_{\bar{p}}$, as actually observed in~\cite{Calore:2022stf}.

A more tractable yet robust definition of the likelihood consists in replacing the distance $\chi_{\rm LiBeB}^{2}$ by the posterior probability $\Pi(\lambda)$ yielded by an independent LiBeB analysis. A Bayesian approach, like in~\cite{Cui:2018nlm}, allows to derive the probability
\begin{equation}
{\cal P}(\mu) \propto {\displaystyle \int} {\cal L}(\bar{p}_{\,} |_{\,} \mu , \lambda) \; \Pi(\lambda) \; d{\lambda} \,,
\end{equation}
which $\bar{p}$ data associate to each set $\mu$ of DM parameters.
From a frequentist perspective, $\chi_{\rm LiBeB}^{2}$ could be replaced by a nuisance term gauging the distance between the CR parameters $\lambda$ and those $\hat{\lambda}$ best-fitting the LiBeB data. It is even possible to go a step further by noticing that the primary DM component of the $\bar{p}$ flux scales as ${L^{2}}/{K}$, i.e. as $L$ once the B/C ratio is taken into account. In~\cite{Calore:2022stf}, the nuisance term that replaces $\chi_{\rm LiBeB}^{2}$ gauges the distance of $L$ from the LiBeB best-fit value $\hat{L}$, hence the likelihood
\begin{equation}
- 2 \ln {\cal L}(L,\mu) =
\left\{ {\displaystyle \frac{\log L - \log \hat{L}}{\sigma_{\log L}}} \right\}^{\!2} +
\chi_{\bar{p}}^{2}(L , \hat{\lambda}_{i} , \mu) \,,
\text{  with $\hat{\lambda}_{i}$ best-fitting LiBeB at fixed $L$.}
\label{eq:definition_ln_cal_L_2}
\end{equation}

A key ingredient in the calculation of the likelihood is the definition of the chi-square. There's the rub! In principle, one should take into account the correlations between different rigidity bins $i$ and $j$, and use a covariance matrix of errors for both data (${\cal C}^{\rm data}$) and theory (${\cal C}^{\rm model}$), hence the definition
\begin{equation}
\chi_{\bar{p}}^{2} \equiv \sum_{i,j} \, x_{i\,} ({\cal C}^{-1})_{ij\,} x_j \,,
\end{equation}
where $x_{i} = \Phi_{\bar{p} , i}^{\rm exp} - \Phi_{\bar{p} , i}^{\rm th}$ while $\cal{C} = \cal{C}^{\rm data} + \cal{C}^{\rm model}$. It is unfortunate that $\cal{C}^{\rm data}$ is not yet published. Its structure can nevertheless be guessed as explained in~\cite{Derome:2019jfs}. The matrix $\cal{C}^{\rm model}$ results from uncertainties in the secondary $\bar{p}$ production cross sections, in the fluxes of the progenitors producing them and in the CR propagation parameters $\lambda_{i}$ other than $L$~\cite{Boudaud:2019efq}.
A close inspection of Table~1 indicates that most of the analyses where a covariant matrix of errors is used do not point toward a DM signal, whereas the opposite conclusion is reached if statistical and systematic experimental errors are just added in quadrature.

As a final word of caution, it is tempting to gauge the statistical significance of the null hypothesis by interpreting the likelihood ratio as a $\Delta \chi^{2}$ associated to the two degrees of freedom $\mC$ and $\sigv$. However, the DM mass is not defined under the null hypothesis. In these conditions, the global significance should be determined by generating mock $\bar{p}$ data under the null hypothesis (only secondaries) to build the $\Delta \chi^{2}$ law and derive the actual $p$-value~\cite{Heisig:2020nse}.

%
\begin{figure}[t!]
\centering
\includegraphics[width=0.49\columnwidth]{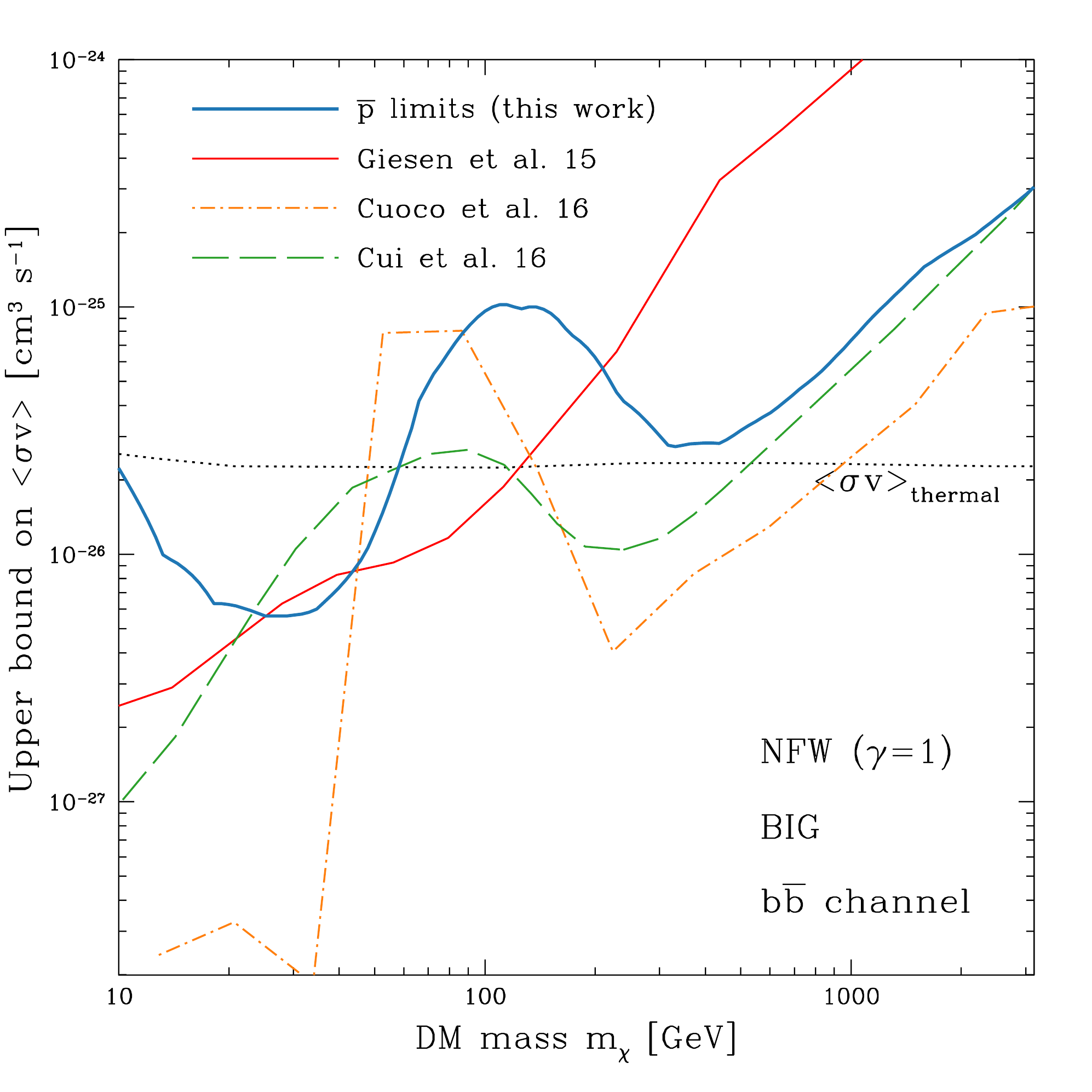}
\includegraphics[width=0.49\columnwidth]{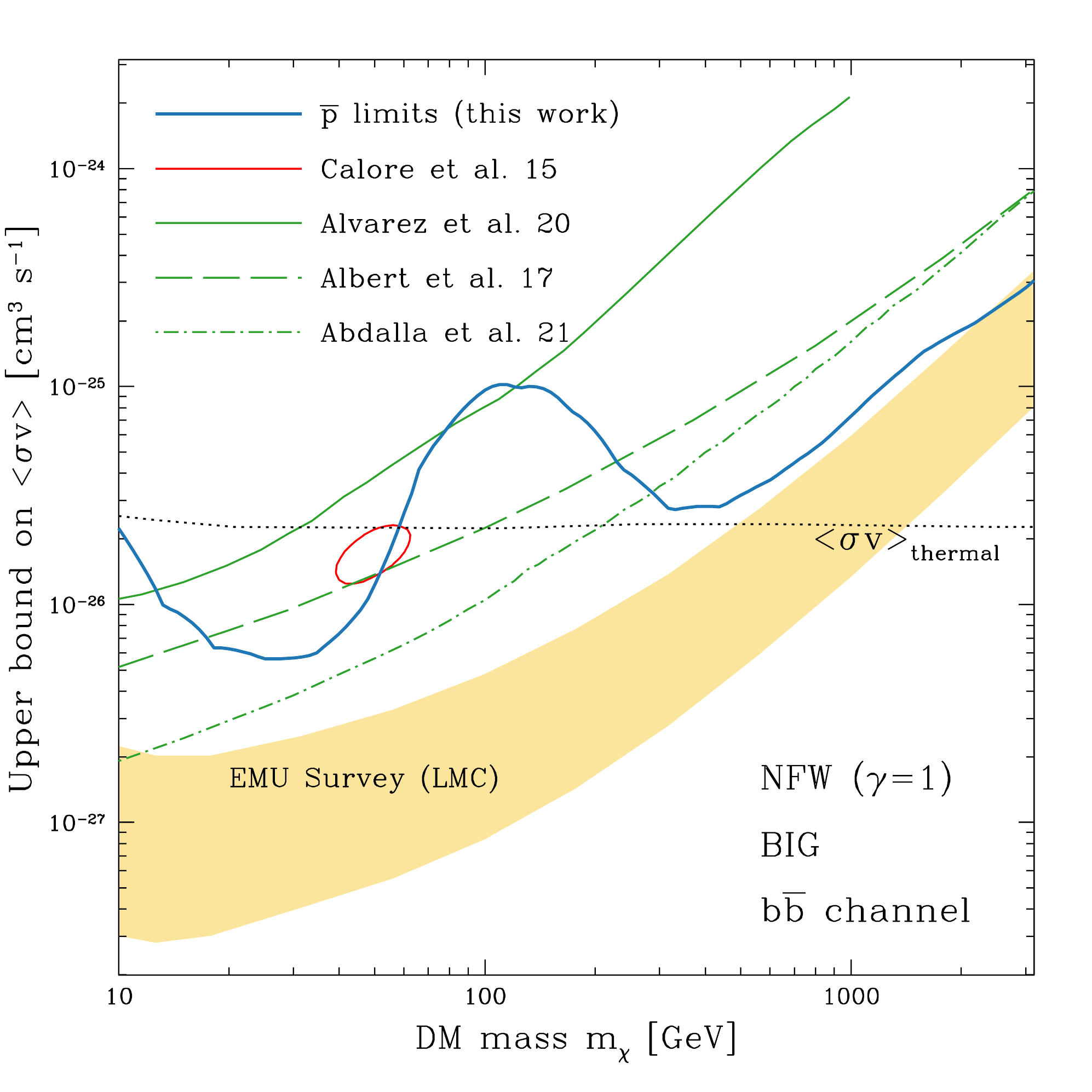}
\caption{
{\it Left panel:}
The upper limit on the annihilation cross section derived in~\cite{Calore:2022stf} for the $b\bar{b}$ channel (blue solid line), compared to other results involving $\bar{p}$ analyses of~\cite{AMS:2016oqu}.
The red solid line corresponds to the Giesen et al. upper bound~\cite{Giesen:2015ufa}. The limits set by Cuoco et al.~\cite{Cuoco:2016eej} (orange dot-dashed curve) and Cui et al.~\cite{Cui:2016ppb} (green dashed line) are also featured.
{\it Right panel:}
The upper limit on the annihilation cross section derived in~\cite{Calore:2022stf} for the $b\bar{b}$ channel (blue solid line), compared to other probes.
The red contour is the 95\% CL contour of the fit to the Galactic Center excess reported in~\cite{Calore:2014xka} for the same annihilation channel. Bounds from different samples of dwarf spheroidal galaxies (dSph) derived with a new data-driven method~\cite{Alvarez:2020cmw} (green solid line), with traditional template-fitting strategies~\cite{Fermi-LAT:2016uux} (green dashed line), and by combining Fermi-LAT and ground-based telescopes data~\cite{Hess:2021cdp} (Glory Duck project, green dot-dashed line) are also displayed. The yellow band are radio constraints obtained from the EMU survey~\cite{Regis:2021glv}. The thermal relic cross section is reported in dotted black lines.}
\label{fig:pbar_lilmits_on_DM}
\end{figure}
%

\section{Anti-nuclei -- The new frontier}
\label{sec:antinuclei}

Although not seen so far, anti-nuclei are also expected in the cosmic radiation. They should be produced as secondary species, like in terrestrial accelerators, by the collisions of CR primaries on Galactic gas.
The General Antiparticle Spectrometer (GAPS) is about to fly and will measure the $\bar{p}$ flux below $200$ MeV~\cite{GAPS:2022ncd}. The instrument is devised to disentangle antiprotons from antideuterons.
The AMS collaboration has recently reported~\cite{Choutko}
a few events in the mass region from 0 to 10 GeV with charge $Z = -2$ and rigidity ${\cal R} < 50$ GV. The masses of all events are in the $^{3}\overline{\rm He}$ and $^{4}\overline{\rm He}$ mass region. The event rate is one $\overline{\rm He}$ for $10^{8}$ ${\rm He}$.
This is extremely surprising insofar as the flux of a secondary anti-nucleus decreases by 4 orders of magnitude each time its atomic number is incremented by 1. The flux of secondary $^{3}\overline{\rm He}$ is expected to be well below AMS-02 sensitivity~\cite{Korsmeier:2017xzj,Poulin:2018wzu}.

These AMS events, if confirmed, would nevertheless point to unconventional physics. In general, DM annihilations or decays yield also a vanishingly small flux. However, a nice counterexample has been recently proposed in~\cite{Winkler:2020ltd}, where the DM particles annihilate into $b{\bar{b}}$ pairs, a fraction ${\cal F} \simeq 0.1$ of which hadronize into $\bar{\Lambda}_{b}$ baryons. This fraction has been measured at LEP but the event generator Pythia is short by a factor of $3$ to reproduce it. The colored string extending between the $b$ and $\bar{b}$ quarks needs to be broken into a diquark ($ud$) and an anti-diquark ($\bar{u}\bar{d}$) pair. In Pythia, the probability for this to happen, dubbed {\tt probQQtoQ}, is too low. It has been increased from 0.09 to 0.24 in~\cite{Winkler:2020ltd} to recover the value of ${\cal F}$ measured at LEP.
Increasing {\tt probQQtoQ} has also the merit to enhance the decay of $\bar{\Lambda}_{b}$ baryons into $^{3}\overline{\rm He}$ nuclei. In the process, two protons are created to conserve the baryon number. Given the masses of the initial ($5.6$ GeV) and final ($4.7$ GeV) states, the nucleons and anti-nucleons are produced at rest and the coalescence of anti-nucleons into $^{3}\overline{\rm He}$ occurs easily, hence the possibility to explain the putative AMS events.
This proposal has been criticized insofar as modifying {\tt probQQtoQ} plays havoc with many observations, leading to an overproduction of baryons with respect to mesons at LEP and the LHC~\cite{Kachelriess:2021vrh}.

\section{Conclusion}
\label{sec:conclusion}

The transport of charged particles inside our Galaxy is better understood, especially in the light of recent AMS-02 measurements of secondary nuclear species. The size of the magnetic halo $L$ is found to be $4.5 \pm 1$ kpc.
Anomalies in the fluxes of antimatter charged CRs could be an indirect signature of annihilating DM particles. But caution must prevail. Positrons, for instance, are most probably accelerated in PWNe and are detected as $\gamma$-ray TeV haloes. As regards antiprotons, many groups have reported an excess. But taking properly the errors into account, i.e. including their correlations at different energies, makes the excess recede.
Finally, AMS-02 has reported a few anti-helium events. If confirmed, this would be a major discovery, pointing to exotic physics. If DM annihilates into $b \bar{b}$ quarks, the decay of $\bar{\Lambda}_{b}$ into $^{3}\overline{\rm He}$ could produce an excess over secondaries provided that diquark formation is enhanced. The search for this reaction at the LHC and the measurement of its branching ratio would definitely be of great interest.

\section*{Acknowledgements}
Many thanks to the organizers for their invitation and for the kind and inspiring atmosphere of the conference. I also would like to thank my home Institution LAPTh for its financial support.





\nolinenumbers
\end{document}